\documentstyle[aps,graphicx,prl]{revtex}

\newcommand{\myscalebox}[1]{\scalebox{0.43}[0.43]{#1}}
   
\begin{document}

\twocolumn[\hsize\textwidth\columnwidth\hsize\csname@twocolumnfalse\endcsname

\title{Typical solution time for a vertex-covering algorithm on 
   finite-connectivity random graphs}
\author{Martin Weigt and Alexander K. Hartmann}
\address{Institute for Theoretical Physics,
      University of G\"ottingen, Bunsenstr. 9, 37073 G\"ottingen, Germany
        }

\date{\today}
\maketitle
 
\begin{abstract}
In this letter, we analytically describe the typical solution time
needed by a backtracking algorithm to solve the vertex-cover problem
on finite-connectivity random graphs. We find two different transitions:
The first one is algorithm-dependent and marks the dynamical
transition from linear to exponential solution times. The second
one gives the maximum computational complexity, and is found exactly 
at the threshold where the system undergoes an algorithm-independent 
phase transition in its solvability. Analytical results are
corroborated by numerical simulations.
\end{abstract}
\vskip.5pc]

\narrowtext
Over the last few years, phase-transition phenomena in combinatorial
problems have increasingly attracted computer scientists and, more
recently, also statistical physicists \cite{review1,review2}.  Many
computationally hard problems, as {\it e.g.} 3-satisfiability, graph
coloring, number partitioning or vertex cover \cite{GaJo}, undergo
dramatic changes in their solvability or their solution structure when
external parameters are changed. These problems, all belonging to the
class of NP-complete problems \cite{GaJo}, are believed to be solvable
only in a time which scales exponentially with the problem
size. Therefore the scientific interest was largely increased by the
observation that phase transitions are strongly related to a
pronounced peak in the typical computational time: The hardest
instances were typically found in the vicinity of the transition
point, where problems are said to be critically constrained. Far away
from this point, problems are easily solved or hopelessly
over-constrained. Problems at such phase boundaries thus provide an
optimal testing ground for the development or improvement of
algorithms.

Classical complexity theory characterizes the hardness of a
computational problem with respect to the worst possible case. The
above-mentioned observations have however underlined the need of a
{\it typical-case complexity theory}. At this point statistical
mechanics enters: Many algorithm-independent aspects of these
phenomena, as {\it e.g.} the location of the phase transition and the
solution space structure, have already been characterized using
methods from statistical mechanics of disordered systems,
\cite{MoZe,nature,Me,WeHa}. A description of the average
behavior of specific algorithms is however less obvious.
Probabilistic methods help to analyze simple descent algorithms and
thus establish rigorous bounds on phase boundaries
\cite{ChFr,review1,review2}, but the calculation of computing times
for complete backtracking algorithms was out of range. Recently a
breakthrough was obtained in \cite{CoMo} for the 3-satisfiability
problem: Combining elements of probabilistic analysis with methods
from statistical mechanics, the typical time complexity of a
backtracking algorithm could be obtained.

{\bf The vertex-cover decision problem:}
In this letter we concentrate on the vertex-cover (VC) problem on
finite-connectivity random graphs, which is one of the basic
NP-complete combinatorial problems, see \cite{GaJo}. It is 
expected that no algorithm can be designed which solves this
problem always in a time growing sub-exponentially with the graph size.
VC was recently shown to have similar phase-transition 
properties as satisfiability, but it is much easier to understand 
due to its simpler geometrical structure \cite{WeHa}. After having
introduced the model and reviewed some recent results, we will
introduce a simple branch-and-bound algorithm and calculate its
computational time complexity by means of analytical as well as
numerical tools.

Vertex covers are defined as follows: Take any undirected graph
$G=(V,E)$ with $N$ vertices $i\in V=\{1,2,...,N\}$ and $M$ edges
$\{i,j\}\in E\subset V\times V$.  We consider subsets $V_{vc}\subset
V$; vertices $i$ with $i\in V_{vc}$ are called covered, and uncovered
for $i\notin V_{vc}$.  Analogously also an edge ${i,j}\in E$ is called 
covered iff at least one of its end-vertices is covered, $i\in V_{vc}$ or
$j\in V_{vc}$.  The set $V_{vc}$ is a {\em vertex cover} iff all edges 
of the graph are covered.

The {\em vertex-cover decision problem} asks whether there are VCs of
fixed given cardinality $xN=|V_{vc}|$. In other words we are
interested if it is possible to cover all edges of $G$ by covering
$xN$ suitably chosen vertices, {\it i.e.}  by distributing $xN$ {\em
covering marks} among the vertices.

In order to be able to speak of typical or average cases, we have to
introduce an ensemble of graphs. We investigate random graphs $G_{N,c/N}$
with $N$ vertices and edges $\{i,j\}$ which are drawn randomly and
independently with probability $c/N$, thus the average
connectivity $c$ remains finite in the large-$N$ limit. For a complete
introduction to the field see \cite{Bo}.

When the number $xN$ of covering marks is lowered ($c$ is kept
constant), the model is expected to undergo a coverable-uncoverable
transition. Using probabilistic tools, rigorous lower and upper bounds
for this threshold \cite{Ga} and the asymptotic behavior for large
connectivities \cite{Fr} have been deduced. Recently we have
investigated VC using a statistical mechanics approach \cite{WeHa}: 
For connectivity $c<e\simeq 2.72$ the transition is given by
\begin{equation}
  \label{eq:xcc}
  x_c(c) = 1- \frac{2W(c)+W(c)^2}{2c}\,,
\end{equation}
where $W(c)$ is the Lambert-$W$-function defined by $W(c)\exp(W(c))=c$.
For $x>x_c(c)$, vertex covers of size $xN$ exist with probability one, 
for $x<x_c(c)$ the available covering marks are not sufficient. For
connectivities $c>e$ replica symmetry breaking is present, and no exact
result for $x_c(c)$ has been obtained.

{\bf The algorithm:} 
We analyze a branch-and-bound algorithm similar to \cite{tarjan77},
for an introduction to this kind of algorithms see \cite{lawler66}. As
each vertex is either covered or uncovered, there are ${N\choose X}$
possible configurations which can be arranged as the leaves of a
binary (backtracking) tree. The basic idea is to traverse the whole
tree to search for vertex covers. At first we explain how the
tree-traversal is organized ({\em branch}) then we show how much
computational time can be saved by excluding subtrees where surely no
covers can be found ({\em bound}).

We introduce three states of vertices: {\em free}, {\em covered} or
{\em uncovered}.  The algorithm starts at the root of the tree where
all vertices are {\em free}. The algorithm descends into the tree by
choosing {\em free} vertices at random. Each vertex $i$ has two
subtrees corresponding to covering/uncovering $i$. If $i$ has
neighboring vertices which are either {\em free} or {\em uncovered},
we mark $i$ {\em covered} first (left subtree).  If the number of
covered vertices does not exceed $xN$ the descent continues. If the
algorithm returns, vertex $i$ is set {\em uncovered} (right
subtree). In case $i$ has only covered neighbors, the order the two
subtrees is exchanged. The algorithm stops either if it has covered
all edges before having used all covering marks (output: graph
coverable) or if its has exhausted all covering marks in the
right-most branch without having covered all edges (output: graph
uncoverable).

The performance of this algorithm can be improved easily by
introducing a {\it bound}. If at any node one of the subtrees can be
proven to contain no VC, the corresponding subtree can be omitted. The
bound used here is simple: It forbids to mark a vertex uncovered if it
has any neighbor which was already marked uncovered. Otherwise
some edges would remain uncovered. The algorithm is summarized below, 
where $G=(V,E)$ denotes the graph, $m(i)\in\{${\em free,cov,uncov}$\}$ 
contains the marks, and $X$ equals the currently available number of 
marks. Initially we set $m(i)=free$ for all $i\in V$, and $X=xN$.

\newlength{\tablen}
\settowidth{\tablen}{xxx}
\newcommand{\tabspace}{\hspace*{\tablen}}
\begin{tabbing}
\tabspace \= \tabspace \= \tabspace \= \tabspace \= \tabspace \=
\tabspace \= \kill
{\bf procedure} vertex-cover($G, m, X$)\\
{\bf begin}\\
\> {\bf if} all edges are covered {\bf then}\\
\>\> {\bf stop};\\
\> {\bf if} $X=0$ {\bf then}\\
\>\> {\bf return};\\
\> Select a vertex $i$ with ($m(i)=$ {\em free}) randomly;\\
\> {\bf if}  $i$ has  neighbors $j$ with $m(j)\neq$ {\em cov} {\bf then}\\
\> {\bf begin}\\
\>\> $m(i)\leftarrow$ {\em cov};\\
\>\> vertex-cover($G, m, X-1$);\\
\>\> {\bf if} $i$ has no neighbors with $m(j)$={\em uncov} {\bf then}\\
\>\> {\bf begin}\\
\>\>\> $m(i)\leftarrow$ {\em uncov};\\
\>\>\> vertex-cover($G, m, X$);\\
\>\> {\bf end}\\
\> {\bf end}\\
\> {\bf else}\quad (all neighbors $j$ of $i$ have $m(j)=$ {\em cov})\\
\> {\bf begin}\\
\>\> $m(i)\leftarrow$ {\em uncov};\\
\>\> vertex-cover($G, m, X$);\\
\>\> $m(i)\leftarrow$ {\em cov};\\
\>\> vertex-cover($G, m, X-1$);\\
\> {\bf end}\\
{\bf end}\\
\end{tabbing}

This algorithm is complete, {\it i.e.} decides whether or not a graph
is coverable with the desired number of covering marks. Due to
backtracking it will in general need exponential time in order to
decide this question. In the following, the solution time is measured
as the {\it number of visited nodes} in the backtracking tree.

{\bf The first descent into the tree:} The analysis of the first
descent into the left-most subtree is straight forward for our
algorithm, as it forms a Markov process of random graphs. In every
time step, one vertex and all its incident edges are covered and can
be regarded as removed from the graph. As the order of appearance of
the vertices is not correlated to its geometrical structure, the graph
remains a random graph.  After $T$ steps, we consequently find a graph
$G_{N-T,c/N}$ having $N-T$ vertices. As the edge probability remains
unchanged, the average connectivity decreases from c to $(1-T/N)c$.

For large $N$, it is reasonable to work with the {\it rescaled time}
$t=T/N$, which becomes continuous in the thermodynamic limit. In this
notation, our generated graph reads $G_{(1-t)N,c/N}$. An isolated
vertex is now found with probability $(1-c/N)^{(1-t)N-1}\simeq
\exp\{-(1-t)c\}$, so the expected number of free covering marks
becomes $X(t)=X- N\int_0^t dt^{'}(1- \exp\{-(1-t^{'})c\})$. The first descent
thus describes a trajectory in the $c-x$-plane,
\begin{eqnarray}
  \label{eq:traj}
  c(t) &=& (1-t)c \\ x(t) &=& \frac{x-t}{1-t} +
\frac{e^{-(1-t)c}-e^{-c}}{(1-t)c} \nonumber
\end{eqnarray}
The results are presented in figure \ref{figTraj}. There are two
cases: for large starting value of $x$, $x(t)$ reaches one at a
certain rescaled time $t<1$, and the graph is proven to be coverable
after having visited $tN$ nodes of the backtracking tree. This holds
as long as the starting point $(x,c)$ is situated above the line
\begin{equation}
  \label{eq:bound}
  x_b(c) = 1+\frac{e^{-c}-1}{c}
\end{equation}
Below $x_b(c)$, $x(t)$ vanishes already before having covered all
edges. So the algorithm has to backtrack, and, intuitively, 
exponential solution times have to be expected.

{\bf The backtracking time:} In order to calculate the solution time
also for $x<x_b(c)$, we combine equations (\ref{eq:xcc}) and
(\ref{eq:traj}).  We have also included $x_c(c)$ into
fig. \ref{figTraj}.  For $x<x_b(c)$, the trajectory of the first
descent crosses the phase transition line at a certain rescaled time
$\tilde{t}$ at $(\tilde{c},\tilde{x})$. There the generated random
subgraph of $\tilde{N}=(1-\tilde{t})N$ vertices and average
connectivity $\tilde{c}$ becomes uncoverable by the remaining
$\tilde{x}\tilde{N}$ covering marks.  Please note that $\tilde{t}=0$
for $x<x_c(c)$, {\it i.e.} if we already start with an uncoverable
graph.  To prove this uncoverability the algorithm has to completely
backtrack the subtree. This part of the algorithm obviously
contributes the exponentially dominating part to the solution time.
In the following we may thus concentrate completely on the generated
subgraph, skipping ``sub-'' in subgraph, subtree, subproblem etc.

Numerical simulation show that the exponential solution times approach
a log-normal distribution of large $N$. Hence, the typical solution time 
$e^{N\tau(x,c)}$ follows from the quenched average
$\tau(x,c)=\lim_{N\to\infty} 1/N \overline{\log(t_{bt}(
G_{\tilde{N},\tilde{c}/\tilde{N}},\tilde{x}))}$ where
$t_{bt}(G_{\tilde{N},\tilde{c}/\tilde{N}},\tilde{x})$ is the
backtracking time for the generated uncoverable instance
$G_{\tilde{N},\tilde{c}/\tilde{N}}$, and the overbar denotes the
average over the random graph ensemble. Solution times, as already
mentioned above, are measured as the number of nodes visited by an
algorithm. Since also the leaves are visited nodes, $t_{bt}$ exceeds 
the number ${\cal{N}}_l$ of leaves. As the depth of the backtracking 
tree is at most $\tilde{N}$, we also have 
$t_{bt}\leq \tilde{N} {\cal{N}}_l$. The exponential time contribution 
is thus given by
\begin{equation}
  \label{eq:entropy}
 \tau(x,c) =\lim_{N\to\infty} \frac{1}{N} \overline{
 \log({\cal{N}}_l (G_{\tilde{N},\tilde{c}/\tilde{N}},\tilde{x}))}\ .
\end{equation}
We have consequently reduced the problem of calculating the
backtracking time to an entropic calculation  which can be achieved
using the tools of equilibrium statistical mechanics.
The number of leaves is trivially bounded 
from above by ${\tilde{N}\choose\tilde{x}\tilde{N}}$,
{\it i.e.} by the number of possible placements of the
$\tilde{x}\tilde{N}$ covering marks on the $\tilde{N}$ vertices. Using 
Stirling´s formula, we thus find
\begin{equation}
  \label{eq:simple}
  \tau(x,c) \leq -\frac{\tilde{c}}{c} \left(\tilde{x}\log\tilde{x}
                  +(1-\tilde{x})\log(1-\tilde{x})\right)\ .
\end{equation}
This time is realized by our algorithm if the bound is skipped, {\it
  i.e.} if all branches of the backtracking tree are visited until
the covering marks are exhausted. Using the bound,
our algorithm does not mark any two neighboring vertices
simultaneously as uncovered. This excludes the most of all
${\tilde{N}\choose\tilde{x}\tilde{N}}$ above-mentioned configurations,
leaving only an exponentially small fraction. So the simple bound causes 
already an exponential speed-up. The number of leaves
fulfilling our criterion can be characterized easily: Imagine a
certain leaf is reached at level $\kappa \tilde{N}$ of the 
backtracking-subtree. Then, our algorithm has constructed a VC of
the subgraph consisting of the $\kappa \tilde{N}$ visited vertices
because edges between these are not allowed to stay uncovered. Due
to the random order of levels in the backtracking tree, this subgraph
is again a random graph $G_{\kappa \tilde{N},\tilde{c}/\tilde{N}}$
having average connectivity $\kappa\tilde{c}$. We may thus conclude
that the number of leaves at level $\kappa \tilde{N}$ equals the
total number ${\cal{N}}_{VC}(G_{\kappa \tilde{N},\tilde{c}/\tilde{N}},
\tilde{x})$ of VCs of 
$G_{\kappa \tilde{N},\tilde{c}/\tilde{N}}$ using 
$\tilde{x}\tilde{N}$ covering marks. Summing
over all possible values of $\kappa$ leads to the
saddle point
\begin{equation}
  \label{eq:time}
  \tau(x,c) =  \mbox{max}_\kappa \left[ 
    \lim_{N\to\infty} \frac{1}{N}\overline{\log 
       {\cal{N}}_{VC}(G_{\kappa \tilde{N},\tilde{c}/\tilde{N}},
        \tilde{x}\tilde{N})
               }\right].
\end{equation}
The average of $\log{\cal{N}}_{VC}$ over the random graph ensemble can
be calculated using the replica trick. In order to avoid
technicalities, we use the annealed bound 
$\overline{\log {\cal{N}}_{VC}} \leq \log \overline{{\cal{N}}_{VC}}$
which provides a very good approximation. The latter average is
calculated easily, we obtain
\begin{eqnarray}
  \label{eq:ann}
  \tau(x,c) &\simeq& \frac{\tilde{c}}{c}
  \mbox{max}_{\kappa=\tilde{x},...,1} \left[ \kappa 
  s_{ann}\left(\frac{\tilde{x}}{\kappa},\tilde{c}\kappa\right)\right]\\ 
  s_{ann}(\tilde{x},\tilde{c}) &=& -\tilde{x} \log \tilde{x} 
    -(1-\tilde{x})\log (1-\tilde{x})-\frac{\tilde{c}}{2}(1-\tilde{x})^2
  \nonumber 
\end{eqnarray}
where $\tilde{x}$ and $\tilde{c}$ follow from the crossing point of
(\ref{eq:traj}) with $x_c(c)$. In fig. \ref{figTime} this result is
compared with numerical simulations. Due to the exponential time
complexity the system sizes which can be treated are of course much 
smaller than for the study of the
first descent.  In order to eliminate however strong logarithmic
finite size dependencies, we have also used the number of leaves in
these simulations; cross-checks using the number of visited nodes in
the backtracking trees also show the expected behavior. Clear
consistency of numerical and analytical data is found. One also finds
that the computational complexity is maximal at $x=x_c(c)$ for both
algorithms with or without bound, as described by equations
(\ref{eq:simple}) and (\ref{eq:ann}).

{\bf Conclusion and outlook:} To conclude, we have calculated the 
typical solution time needed by a complete backtracking algorithm 
for vertex covering random graphs. We have combined probabilistic
methods used in computer science for characterizing the first descent,
and statistical mechanics methods which enabled us to calculate the
phase transition threshold and the entropy of leaves. 

These results imply a very intuitive picture for the different
regimes of the typical computational complexity which is expected to
share essential features with the behavior of
more complicated algorithms. The algorithm starts its first descent
into the backtracking tree. There is some parameter range of the model
inside the coverable phase, where the first descent already successfully
produces a VC and thus proves the coverability of the graph with the
prescribed number of covering marks. In this region the solution time
is found to be typically linear in problem size. If we lower the
allowed number of covering marks, the initial problem still remains 
coverable but the first descent into the tree generates an uncoverable
macroscopic subproblem. To escape from the corresponding subtree, the
algorithm has to backtrack and consequently consumes exponential time.
The maximum backtracking tree appears when the initial problem is
exactly situated at the phase transition point $x=x_c(c)$, there the
exponential solution time shows its maximum, as found also for other
algorithms \cite{WeHa} or other combinatorial problems
\cite{review1,review2}. The height of the time peak depends however on 
the considered algorithm, and consequently also the maximal analyzable
system size. In \cite{WeHa}, we could numerically solve systems up to
$N=140$, but the analysis of this algorithm goes far beyond the 
presented methods.

Related to the depicted scenario, there are mainly two different 
possibilities of improving  algorithms: 

(i) More sophisticated heuristics for the first descent allow
to shift the onset of exponential complexity towards $x_c(c)$. One
important restriction to obtain algorithms analyzable within 
the described scheme is that the order of appearance of
vertices in the backtracking tree must be independent of the
structure of the graph ({\it e.g.} of the connectivities) in order to
remain inside the random graph ensemble. 

(ii) The second possibility of improving algorithms is given by the 
inclusion of more elaborated 
bounds into the backtracking tree. These result in
an exponential speed-up of the algorithm, as we have already seen for
the simple bound used in our algorithm.

{\bf Acknowledgements:} We are very grateful to S. Cocco and
R. Monasson for communicating the results of \cite{CoMo} prior to
publication. J. Berg is acknowledged for carefully reading the
manuscript. AKH is financially supported by the DFG ({\em Deutsche 
Forschungsgemeinschaft}) under grant Zi209/6-1.

\newcommand{\captionTraj}
{Trajectories of the first descent in the $(c,x)$ plane. The full
  lines represent the analytical curves, the symbols
  numerical results of one random graph with $10^6$ vertices,
  $c=2.0$ and $x=0.8$, $0.7$, $0.6$, $0.5$ and $0.3$. The trajectories
  follow the sense of the arrows. The dotted line $x_b(c)$
  separates the regions where this simple algorithm finds a cover from
  the region where the method fails. No trajectory crosses this
  line.  The long dashed line represents the true phase boundary
  $x_c(c)$, instances below that line are not coverable.}

\newcommand{\captionTime} {Normalized and averaged logarithm of running
  time of the algorithm as a function of the fraction $x$ of coverable
  vertices. The solid line is the result of the annealed calculation. 
  The symbols represent the numerical data for $N=12,25,50$, lines are 
  guide to the eye only.  
}

\begin{figure}[htb]
\begin{center}
\myscalebox{\includegraphics{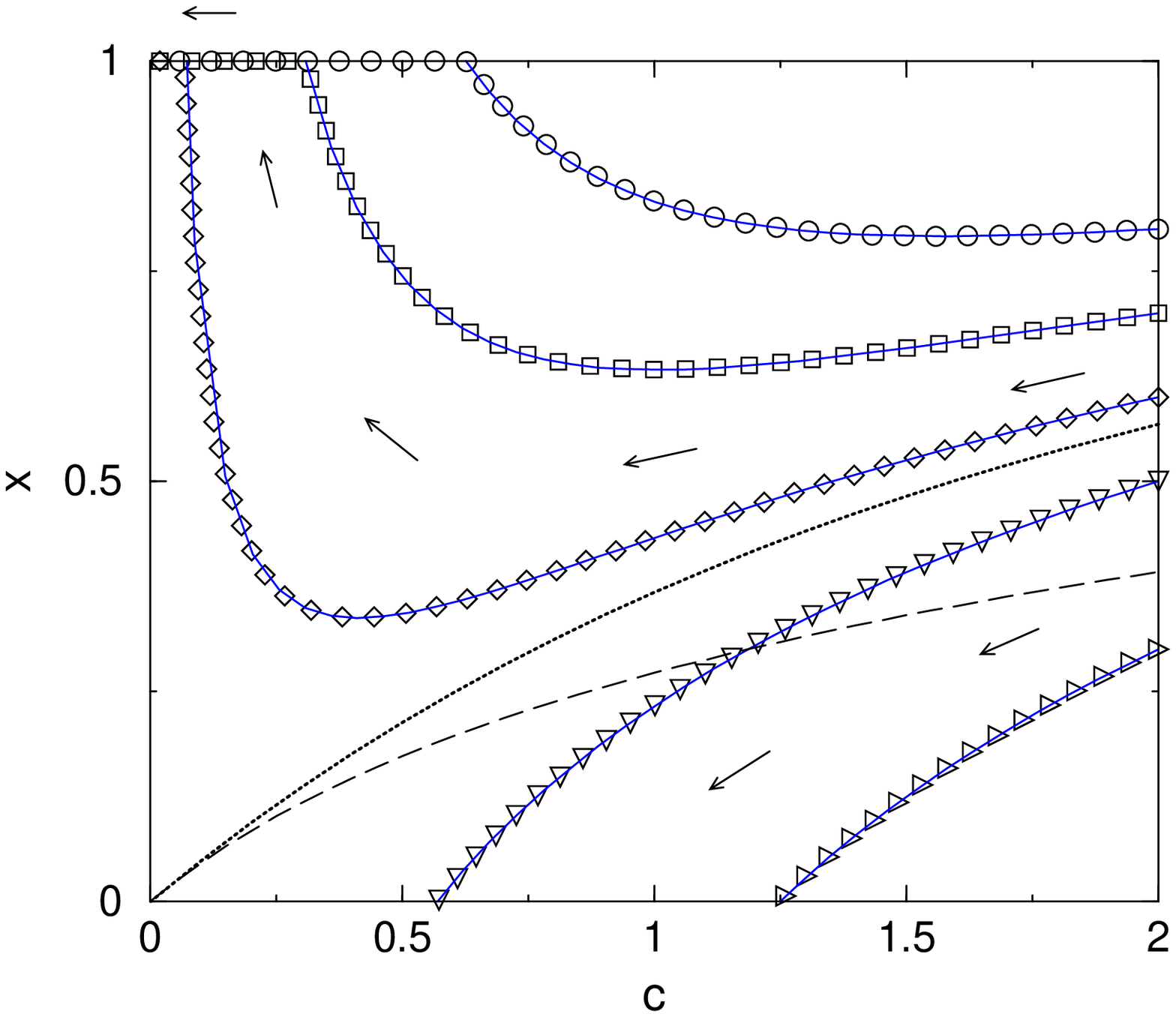}}
\end{center}
\caption{\captionTraj}
\label{figTraj}
\end{figure}

\begin{figure}[htb]
\begin{center}
\myscalebox{\includegraphics{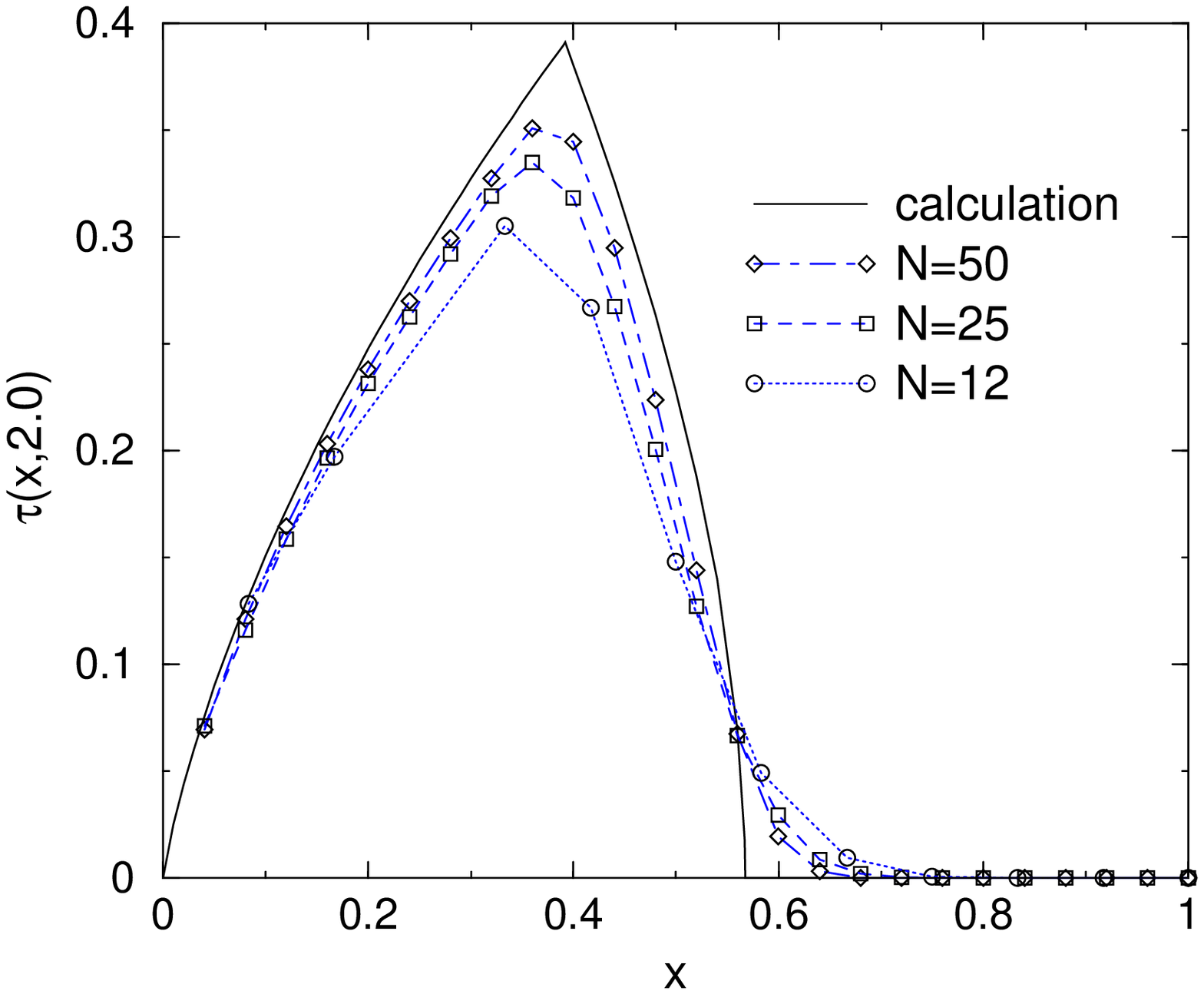}}
\end{center}
\caption{\captionTime}
\label{figTime}
\end{figure}

\end{document}